\documentclass[a4paper]{jpconf}
\usepackage{graphicx}
\usepackage{amsmath}
\usepackage{bm}
\usepackage{dcolumn}
\usepackage{longtable}
\newcolumntype{.}{D{x}{}{-1}}
\newcommand{\bsigma}{\vec{\sigma}}
\newcommand{\cE}{{\cal E}}
\newcommand{\lbr}{\left<} \newcommand{\rbr}{\right>}
\newcommand{\bfp}{\vec{p}}
\newcommand{\bfr}{\vec{r}}
\newcommand{\bnabla}{\vec{\nabla}}
\newcommand{\al}{\alpha}
\newcommand{\Za}{{Z\alpha}}

\begin{document}
\title{Helium fine structure theory for determination of $\bm{\alpha}$}

\author{Krzysztof Pachucki}

\address{Institute of Theoretical Physics, University of Warsaw, Ho\.{z}a 69, 00--681 Warsaw, Poland}

\ead{krp@fuw.edu.pl}

\author{Vladimir A. Yerokhin} \address{
Center for Advanced Studies,
St.~Petersburg State Polytechnical University, Polytekhnicheskaya 29,
St.~Petersburg 195251, Russia }

\ead{yerokhin@pcqnt1.phys.spbu.ru}

\begin{abstract}
We present recent progress in the calculation of the helium fine-structure
splitting of the $2^3P_J$ states, based on the quantum electrodynamic theory.
Apart from the complete evaluation of $m\,\alpha^7$ and $m^2/M\,\alpha^6$ 
corrections, we have performed extensive tests by comparison with
all experimental results for light helium-like ions and 
with the known large nuclear charge asymptotics of individual
corrections. Our theoretical predictions are still limited by the unknown
$m\,\alpha^8$ term, which is conservatively estimated to be  1.7 kHz. However,
comparison with the latest experimental result for the $2^3P_0 - 2^3P_2$ transition
[M.~Smiciklas and T.~Shiner, Phys.~Rev.~Lett. {\bf 105}, 123001 (2010)]
suggests that the higher-order contribution is in fact much smaller than the
theoretical estimate. This means
that the spectroscopic determination of $\alpha$ can be significantly improved
if another measurement of the $2^3P_0 - 2^3P_2$ transition
in helium-like Li$^+$ or Be$^{2+}$ ion is performed.
\end{abstract}

\section{Introduction}
The quantum electrodynamic (QED) theory of atomic energy levels
has achieved a precision level that makes possible the determination of
nuclear properties, like the charge radius, the magnetic dipole, or even 
the nuclear polarizability from measured atomic spectra. If the nuclear
structure effects are negligible or can be eliminated,
one may obtain fundamental constants from comparison of theoretical predictions
with experimental results. The most important examples include
the Rydberg constant determined from hydrogen spectroscopy, the electron
mass derived from the bound-electron
g factor in hydrogen-like ions, and $\alpha$ obtained from the helium fine
structure. 
As first pointed out by Schwartz in 1964 \cite{schwartz:64},
the splitting of the $2^3P_J$ levels in helium can be used for an accurate
determination of the fine structure constant $\alpha$.  The attractive features of
the fine structure in helium as compared to other atomic transitions
are, first, the long lifetime of the metastable $2^3P_J$ levels (roughly two
orders of magnitude longer than that of the $2p$ state in hydrogen) and,
second, the relative simplicity of the theory.
Schwartz's suggestion stimulated a sequence of calculations [2--5],
which resulted in a
theoretical description of the helium fine structure complete up to order
$m\alpha^6$ (or $\alpha^4$~Ry) and a value of $\alpha$ accurate to
0.9~ppm \cite{lewis:78}.

The present experimental precision for the fine-structure intervals in helium is
sufficient for a determination of $\alpha$ with an accuracy of 14~ppb from
Refs.~\cite{zelevinsky:05,borbely:09} and even 5~ppb from Ref.~\cite{shiner:10}. 
In order to match this level of accuracy in the 
theoretical description of the fine structure, the complete calculation of the
next-order, $m\alpha^7$ contribution and an estimation of the higher-order
effects is needed. Work towards this end started in the 1990s and
extended over two decades
[10--19].
In 2006 the
first complete evaluation of the $m\alpha^7$ correction to the helium fine
structure was reported by one of us (KP) \cite{pachucki:06:prl:he}. However, the
numerical results  presented there were in disagreement
with the experimental values by more than 10 standard deviations ($\sigma$).

In our recent investigations \cite{pachucki:09:hefs,pachucki:10:hefs}, we
recalculated, using formulae from Ref. \cite{pachucki:06:prl:he}, all effects 
up to order $m\alpha^7$ to the fine structure of helium
and performed calculations for helium-like ions with nuclear
charges $Z$ up to 10. The calculations were extensively checked by studying
the hydrogenic ($Z\to\infty$) limit of individual corrections
and by comparing them with the results known from the hydrogen theory.
We found several problems in previous numerical calculations and, in
the meantime, the experimental value of the $2^3P_1 - 2^3P_2$ transition
was changed by 3$\sigma$ \cite{borbely:09}. As a result,
the present theoretical predictions are in agreement with the latest
experimental data for the fine-structure intervals in helium, as well as with
most of the experimental data available for light helium-like ions.
Our calculation of the $m\alpha^7$ correction for the fine-structure splitting
in light helium-like atoms was reported in  Refs. \cite{pachucki:10:hefs,pach:10}.
In this paper, we present a detailed description of
all corrections to helium fine structure and a summary of the numerical results.

\section{QED theory of the helium fine structure}

According to the quantum electrodynamic theory (QED)
the energy levels of an atomic system are a function of 
the fine structure constant $\alpha$ and the electron-nucleus mass ratio.
We omit possible nuclear structure effects,
as their contribution to the helium fine structure is negligible.
The fine-structure splitting $E_{\rm fs}(\alpha)$ can be expanded in powers of
$\alpha$,
\begin{equation}
E_{\rm fs} = E_{\rm fs}^{(4)} + E_{\rm fs}^{(5)} + E_{\rm fs}^{(6)}+ E_{\rm fs}^{(7)} + O(\alpha^8)\,.
\end{equation}
The expansion terms  $E_{\rm fs}^{(n)} \equiv m\,\alpha^n\,{\cal E}^{(n)}$ are
of order $m\,\alpha^n$. They implicitly depend on the electron-nucleus mass ratio
and may additionally involve powers of $\ln\alpha$. 
The advantage of this approach is that each of the expansion
terms is expressed as the expectation value of some effective Hamiltonian,
as presented in the following. For convenience, we first consider the infinite nuclear mass
limit, and then account for the finite nuclear mass corrections separately. 

The dominant contribution to the helium fine structure is induced by the
spin-dependent part of the Breit-Pauli Hamiltonian, 
which is, for an infinitely heavy nucleus,
\begin{eqnarray} \label{fs}
H_{\rm fs} & = &\frac{1}{4}\left(
\frac{\bsigma_1\cdot\bsigma_2}{r^3}
-3\,\frac{\bsigma_1\cdot\bfr\,
\bsigma_2\cdot\bfr}{r^5}\right)(1+a_e)^2\, \nonumber \\
& + & \frac{Z}{4} 
\left[
\frac{1}{r_1^3}\,\bfr_1\times\bfp_1\cdot\bsigma_1+
\frac{1}{r_2^3}\,\bfr_2\times\bfp_2\cdot\bsigma_2
\right](1+2a_e)  
\nonumber \\
& + & 
\frac{1}{4\,r^3}\biggl\{
\bigl[(1+2\,a_e)\,\bsigma_2+2\,(1+a_e)\,\bsigma_1\bigr]\cdot\bfr\times\bfp_2
\nonumber \\ &&
-\bigl[(1+2\,a_e)\,\bsigma_1+2\,(1+a_e)\,\bsigma_2\bigr]\cdot\bfr
\times\bfp_1\biggr\}\,,
\label{HSD}
\end{eqnarray}
where  $\bfr = \bfr_1-\bfr_2$. The above Hamiltonian includes
the effect of the anomalous magnetic moment
(amm) $a_e$, which is given by  \cite{kinoshita:07}
(neglecting small vacuum-polarization corrections coming from particles heavier
than an electron)
\begin{eqnarray}
a_e &=& \frac{\alpha}{2\pi} -0.328\,478\,966\,\left(\frac{\alpha}{\pi}\right)^2
  +1.181\,241\,457\,\left(\frac{\alpha}{\pi}\right)^3
  -1.914\,4(35)\,\left(\frac{\alpha}{\pi}\right)^4 +\ldots\,.
\end{eqnarray} 
Expanding the amm prefactors in Eq.~(\ref{fs}), $H_{\rm fs}$ can be written as a
sum of operators contributing to different orders in $\alpha$
\begin{eqnarray} \label{fs2}
H_{\rm fs} = H_{\rm fs}^{(4)}+ \alpha\,H_{\rm fs}^{(5)}+ \alpha^2\,H_{\rm fs,amm}^{(6)}
+ \alpha^3\,H_{\rm fs,amm}^{(7)}+ \ldots\,.
\end{eqnarray} 
Here, $H_{\rm fs}^{(4)}$ and $H_{\rm fs}^{(5)}$ are the complete 
effective Hamiltonians to order $m\,\alpha^4$ and $m\,\alpha^5$, respectively,
whereas $H_{\rm fs,amm}^{(6)}$ and $H_{\rm fs,amm}^{(7)}$ are the amm parts of the
corresponding higher-order operators. The contributions to the fine structure are
\begin{eqnarray}
{\cal E}^{(4)} &=& \bigl< H^{(4)}_{\rm fs} \bigr> + O(m/M)\,,\\
{\cal E}^{(5)} &=& \bigl< H^{(5)}_{\rm fs} \bigr> + O(m/M)\,,
\end{eqnarray}
where the expectation values are calculated with  
the corresponding eigenstate of the nonrelativistic Hamiltonian $H_0$
\begin{equation}
H_0 = \frac{p_1^2+p_2^2}{2}-\frac{Z}{r_1}-\frac{Z}{r_1}+\frac{1}{r}\,.
\end{equation}

The finite nuclear mass corrections up to order $m\alpha^5$ are 
conveniently divided into three parts, termed the mass scaling, the
mass polarization, and the recoil operators. 
The effect of the mass scaling is accounted
for by including the prefactor $(m_r/m)^3$ in the operator $H_{\rm fs}$,
where $m_r$ is the reduced mass for the electron-nucleus system. The effect of
the mass polarization can be accounted for to all orders by evaluating
expectation values of all operators on the 
eigenfunctions of the Schr\"odinger Hamiltonian with the mass-polarization
operator $(m_r/M)\, \bfp_1\cdot\bfp_2$ included. The third effect is
induced by the recoil addition to the Breit-Pauli Hamiltonian 
\begin{eqnarray} \label{fsrec}
H_{\rm fs,rec} &=& \frac{Z}{2}\,\frac{m}{M}\, \left[
\frac{\bfr_1}{r_1^3}\times(\bfp_1+\bfp_2)\cdot\bsigma_1
+ \frac{\bfr_2}{r_2^3}\times(\bfp_1+\bfp_2)\cdot\bsigma_2 
\right]
(1+a_e)\,.
\end{eqnarray}

\section{The spin-dependent $\bm{m}\, \bm{\alpha}^{\bf 6}$ contribution}
\label{sec2}

The $m\,\alpha^6$ contribution to the helium fine structure is a sum of the
second-order perturbation corrections induced by the Breit-Pauli Hamiltonian
and the expectation value of 
the effective fine-structure Hamiltonian to this order, $H^{(6)}_{\rm fs}$, 
\begin{eqnarray}  \label{fs6}
{\cal E}^{(6)} &=& \lbr H_{\rm fs}^{(4)} \frac1{(E_0-H_0)'} H_{\rm fs}^{(4)} \rbr
   + 2\,\lbr H^{(4)}_{\rm nfs} \frac1{(E_0-H_0)'} H_{\rm fs}^{(4)} \rbr
    + \lbr H^{(6)}_{\rm fs} + H^{(6)}_{\rm fs, amm}\rbr\,.
\end{eqnarray} 
Here, $1/(E_0-H_0)'$ is the reduced Green function and
$H^{(4)}_{\rm nfs}$ is the spin-independent part of the Breit-Pauli
Hamiltonian, 
\begin{eqnarray} \label{nfs}
H^{(4)}_{\rm nfs} & = & -\frac{1}{8}\,(p_1^4+p_2^4)+
\frac{Z\,\pi}{2}\,\bigl[\delta^3(r_1)+\delta^3(r_2)\bigr]
-\frac{1}{2}\,p_1^i\,
\biggl(\frac{\delta^{ij}}{r}+\frac{r^i\,r^j}{r^3}\biggr)\,p_2^j
\,,
\end{eqnarray}
where we have omitted a term with $\delta^3(r)$ since it vanishes for
the triplet states.
$H^{(6)}_{\rm fs}$ consists of 15 operators first derived by Douglas and Kroll (DK)
\cite{douglas:74} in the framework of the Salpeter equation. These operators
were later rederived using the much simpler effective field method
in Ref.~\cite{pachucki:99:jpb}. The result is
\begin{equation} 
   H^{(6)}_{\rm fs}  = \sum_{i=1}^{15} B_i\,,
\end{equation} 
where the $B_i$ are given in Table 1.

\begin{table}
\caption{Effective operators contributing to $H^{(6)}_{\rm fs}$ (left column) and $H_H$ (right column)}
\label{table1}
\begin{center}
\(
\begin{array}{ll}\br
\mbox{\rm Operator $\times m\,\alpha^6$} & \mbox{\rm Operator $\times m\,\alpha^7/\pi$} \\
\hline \\[-2ex]
B_1   =-\frac{3\,Z}{8}\,p_1^2\,\frac{\bfr_1}{r_1^3}\times\bfp_1 \cdot \bsigma_1
& H_1 = -\frac{Z}{4}\,p_1^2\,\frac{\vec r_1}{r_1^3}\times\vec p_1\cdot\vec\sigma_1   \\[1ex]
B_2 = -Z\,\frac{\bfr_1}{r_1^3}\times\frac{\bfr}{r^3}\cdot \bsigma_1\,(\bfr\cdot\bfp_2) 
& H_2 = -\frac{3\,Z}{4}\,\frac{\vec r_1}{r_1^3}\times\frac{\vec r}{r^3}
\cdot\vec\sigma_1\,(\vec r\cdot\vec p_2)    \\[1ex]
B_3 =\frac{Z}{2}\,\frac{\vec r}{r^3}\cdot\vec\sigma_1\,
\frac{\vec r_1}{r_1^3}\cdot\vec\sigma_2
& H_3 = \frac{3\,Z}{4}\,\frac{\vec r}{r^3}\cdot\vec\sigma_1\,
\frac{\vec r_1}{r_1^3}\cdot\vec\sigma_2    \\[1ex]
B_4 = \frac{1}{2\,r^4}\,\vec r\times\vec p_2\cdot\vec\sigma_1
& H_4 = \frac{1}{2\,r^4}\,\vec r\times\vec p_2\cdot\vec\sigma_1  \\[1ex]
B_5  = -\frac{1}{2\,r^6}\,\vec r\cdot\vec\sigma_1\,\vec r\cdot\vec\sigma_2
& H_5 = -\frac{3}{4\,r^6}\,\vec r\cdot\vec\sigma_1\,\vec r\cdot\vec\sigma_2   \\[1ex]
B_6   = \frac{5}{8}\,p_1^2\,\frac{\vec r}{r^3}\times\vec p_1\cdot\vec\sigma_1
& H_6 = \frac{1}{4}\,p_1^2\,\frac{\vec r}{r^3}\times\vec p_1\cdot\vec\sigma_1   \\[1ex]
B_7  = -\frac{3}{4}\,p_1^2\,\frac{\vec r}{r^3}\times\vec p_2\cdot\vec\sigma_1
& H_7 = -\frac{1}{4}\,p_1^2\,\frac{\vec r}{r^3}\times\vec p_2\cdot\vec\sigma_1  \\[1ex]
B_8   = -\frac{i}{4}\,p_1^2\,\frac{1}{r}\,\bsigma_1\cdot(\bfp_1\times\bfp_2)
& H_8 = -\frac{Z}{4\,r}\,\frac{\vec r_1}{r_1^3}\times\vec p_2\cdot\vec\sigma_1    \\[1ex]
B_9   =-\frac{3\,i}{4}\,p_1^2\,\frac{1}{r^3}\,\bfr\cdot\bfp_2\,\bfr\times\bfp_1\cdot\bsigma_1
& H_9 = -\frac{i}{2}\,p_1^2\,\frac{1}{r^3}\,\vec r\cdot\vec p_2\,\vec r\times\vec p_1\cdot\vec\sigma_1   \\[1ex]
B_{10} =\frac{3\,i}{8\,r^5}\,\vec r\times (\vec r\cdot \vec p_2)\,\vec p_1\cdot \vec\sigma_1
& H_{10} = \frac{3\,i}{4\,r^5}\,\vec r\times (\vec r\cdot \vec p_2)\,\vec p_1\cdot \vec\sigma_1      \\[1ex]
B_{11} = -\frac{3}{16\,r^5}\,\vec r\times (\vec r\times \vec p_1\cdot\vec\sigma_1)\,\vec p_2\cdot \vec\sigma_2
& H_{11} = -\frac{3}{8\,r^5}\,\vec r\times (\vec r\times \vec p_1\cdot\vec\sigma_1)\,\vec p_2\cdot \vec\sigma_2     \\[1ex]
B_{12} = -\frac{1}{16\,r^3}\,\vec p_1\cdot\vec\sigma_2\,\vec p_2\cdot\vec\sigma_1
& H_{12} = -\frac{1}{8\,r^3}\,\vec p_1\cdot\vec\sigma_2\,\vec p_2\cdot\vec\sigma_1   \\[1ex]
B_{13} = \frac{3}{2}\,p_1^2\,\frac{1}{r^5}\,\vec r\cdot\vec\sigma_1\,\vec r\cdot\vec\sigma_2
& H_{13} = \frac{21}{16}\,p_1^2\,\frac{1}{r^5}\,\vec r\cdot\vec\sigma_1\,\vec r\cdot\vec\sigma_2    \\[1ex]
B_{14} =-\frac{i}{4}\,p_1^2\,\frac{\vec r}{r^3}\cdot\vec\sigma_1\,\vec p_1\cdot\vec\sigma_2
& H_{14} = -\frac{3\,i}{8}\,p_1^2\,\frac{\vec r}{r^3}\cdot\vec\sigma_1\,\vec p_1\cdot\vec\sigma_2   \\[1ex]
B_{15} =\frac{i}{8}\,p_1^2\,\frac{\vec r}{r^3}\cdot\vec\sigma_1\,\vec p_2\cdot\vec\sigma_2
& H_{15} = \frac{i}{8}\,p_1^2\,\frac{1}{r^3}\,\bigl(\vec r\cdot\vec\sigma_2\,\vec p_2\cdot\vec\sigma_1
         +\vec r\cdot\vec \sigma_1\,
         \vec p_2\cdot\vec\sigma_2 
\\[1ex] & \hspace*{7ex}
         -\frac{3}{r^2}\,\vec r\cdot\vec\sigma_1\,\vec
          r\cdot\vec\sigma_2\,\vec r\cdot\vec p_2\bigr)     \\[1ex]
& H_{16} = -\frac{1}{4}\,\vec p_1\cdot\vec \sigma_1\,\vec p_1\times\frac{\vec r}{r^3}\cdot\vec p_2   \\[1ex]
& H_{17} = \frac{1}{8}\,\vec p_1\cdot\vec\sigma_1\,\bigl(-\vec
p_1\cdot\vec\sigma_2\,\frac{1}{r^3}+3\vec p_1\cdot\vec r\,\frac{\vec
  r}{r^5}\cdot\vec\sigma_2\bigr)\\[1ex] \br
\end{array}
\)
\end{center}
\end{table}

The finite nuclear mass corrections to the $m\,\alpha^6$
contribution can be divided into the mass scaling, the mass polarization,
and the operator parts. The mass scaling prefactor is $(m_r/M)^4$ for the 
$B_2$, $B_3$, $B_4$, and $B_5$, $(m_r/M)^5$ for the other $B_i$ operators, 
$(m_r/M)^6$ for the second-order corrections involving the first term in
Eq.~(\ref{nfs}), and $(m_r/M)^5$ for all other second-order corrections. The
mass polarization effect is most easily accounted for by including the mass
polarization operator in the zeroth-order Hamiltonian. The operator part 
comes from recoil corrections to $H^{(4)}_{\rm fs}$, $H^{(4)}_{\rm nfs}$,
and $H^{(6)}_{\rm fs}$. The recoil part of $H^{(4)}_{\rm fs}$ is given by
Eq.~(\ref{fsrec}). The spin-independent recoil part of the  Breit-Pauli
Hamiltonian is 
\begin{align}
  H_{\rm nfs,rec}^{(4)} &\ = 
    -\frac{Z}{2}\,\frac{m}{M} \sum_{a=1,2}
    p_a^i\,\left(\frac{\delta^{ij}}{r_a}+\frac{r^i_ar^j_a}{r_a^3}\right)
  (p_1^j+p_2^j)\,.
\end{align}
Recoil corrections to the DK operators were studied by Zhang \cite{zhang:97}
and by Pachucki and Sapirstein~\cite{pachucki:03:jpb}. The result is given by the
effective Hamiltonian  
\begin{eqnarray} 
  H^{(6)}_{\rm fs,rec}  &=& \frac{m}{M}\,\biggl[
 {i Z \over 4 } p_1^2 {1 \over r_1} \bsigma_1 \cdot (\bfp_1 \times \bfp_2) 
 -\frac{i Z}{4}\,p_1^2\,\frac{\bfr_1}{r_1^3}\,
(\bsigma_1 \cdot \bfr_1\times \bfp_1)\cdot (\bfp_1+\bfp_2) \nonumber\\&&
-\frac{3 Z}{4}\,p_1^2\,\bsigma_1 \cdot \frac{\bfr_1}{r_1^3}\times (\bfp_1+\bfp_2) 
+Z \,\bsigma_1 \cdot \frac{\bfr}{r_1\,r^3}\times (\bfp_1+\bfp_2) 
+Z\, \bsigma_1 \cdot {\bfr \over r^3} \times{\bfr_{1} \over r_{1}^3}\,(\bfr_1 \cdot(\bfp_1 + \bfp_2))
\nonumber \\ &&
+Z^2\,\bsigma_1 \cdot {\bfr_1 \over r_1^3} \times {\bfr_2 \over r_2^3}\,(\bfr_1 \cdot \bfp_1) 
-\frac{Z^2}{2} \,\bsigma_1 \cdot \frac{\bfr_{1}}{r_{1}^4}\times (\bfp_1+\bfp_2) 
-{Z^2 \over 4} \bsigma_1 \cdot {\bfr_2 \over r_2^3} \,
              \bsigma_2 \cdot {\bfr_1 \over r_1^3}\biggr].
\end{eqnarray}

\section{The spin-dependent $\bm{m}\bm{\alpha}^{\bm{7}}$ correction}

The $m\alpha^7$ correction to the helium fine structure
can be conveniently separated into four parts
\begin{align} \label{ma7}
\cE^{(7)} = \cE^{(7)}_{\rm log}+ \cE^{(7)}_{\rm first} + \cE^{(7)}_{\rm sec}+ \cE^{(7)}_{L}\,.
\end{align}
The first term above combines all terms with $\ln Z$ and $\ln \alpha$
[11--13,15,20],
\begin{align} \label{E7log}
\cE^{(7)}_{\rm log} &\ =
\ln[(Z\,\alpha)^{-2}] \,\left[
\lbr \frac{2 Z}{3}\,
i\,\bfp_1\times\delta^3(r_1)\,\bfp_1\cdot\bsigma_1 \rbr
  \right.
-\lbr \frac{1}{4}
   (\bsigma_1\cdot\bnabla)\,(\bsigma_2\cdot\bnabla) \delta^3(r)\rbr
   \nonumber \\ &
-\lbr \frac{3}{2}\,
i\,\bfp_1\times\delta^3(r)\,\bfp_1\cdot\bsigma_1\rbr
   \left.
 + \frac{8Z}{3} \lbr H^{(4)}_{\rm fs} \frac1{(E_0-H_0)'} \bigl[
 \delta^3(r_1)+\delta^3(r_2)\bigr]\rbr
\right] \,.
\end{align}

The second part of $\cE^{(7)}$
is induced by effective Hamiltonians to order $m\alpha^7$.
They were derived by one of us
(K.P.) in Refs. \cite{pachucki:06:prl:he,pachucki:09:hefs}.
(The previous derivation of this correction
by Zhang \cite{zhang:96:a,zhang:96:b} turned out to be not entirely consistent.)
The result is
\begin{align}
\cE^{(7)}_{\rm first} = \Bigl<  H_Q + H_H + H^{(7)}_{\rm fs, amm} \Bigr>\,. \label{16}
\end{align}
The Hamiltonian $H_Q$ is induced by the two-photon exchange
between the electrons, the electron self-energy and the vacuum
polarization. It is given by
\cite{pachucki:06:prl:he}
\begin{align} \label{HQ}
H_Q  &\ = Z\,\frac{91}{180}
\,i\,\bfp_1\times\delta^3(r_1)\,\bfp_1\cdot\bsigma_1
-\frac12\, (\bsigma_1\cdot\bnabla)\,(\bsigma_2\cdot\bnabla)\, \delta^3(r)\,
 \left[ \frac{83}{30}+\ln Z\right]
 \nonumber \\ &
+3\,i\,\bfp_1\times\delta^3(r)\,\bfp_1\cdot\bsigma_1\,
  \left[ \frac{23}{10}-\ln Z\right]
-\frac{15}{8\,\pi}\, \frac{1}{r^7}\,(\bsigma_1\cdot\bfr)\,(\bsigma_2\cdot\bfr)
-\frac{3}{4\,\pi}\,i\,\bfp_1\times \frac{1}{r^3}\, \bfp_1\cdot\bsigma_1\,.
\end{align}
Here, the terms with $\ln Z$ compensate
the logarithmic dependence implicitly present in
the expectation values of the singular operators $1/r^3$ and $1/r^5$, so that
matrix elements of $H_Q$ do not have any logarithms in their $1/Z$ expansion.
The singular operators are defined through their integrals with
the arbitrary smooth function $f$
\begin{align} \label{def1}
\int d^3 r \frac{1}{r^3}\,f(\bfr) \equiv \lim_{\epsilon\rightarrow 0} \int
\,& d^3 r\,\biggl[\frac{1}{r^3}\,\theta(r-\epsilon)
+ 4\,\pi\,\delta^3(r)\,(\gamma+\ln\epsilon)\biggr]\,f(\bfr)
\end{align}
and
\begin{align} \label{def2}
\int d^3 r\,\frac{1}{r^7}&\,\biggl(r^i\,r^j-\frac{\delta^{ij}}{3}\,r^2\biggr)\,f(\bfr) \equiv\\ &
\lim_{\epsilon\rightarrow 0} \int
d^3 r\,
\biggl[\frac{1}{r^7}\,\biggl(r^i\,r^j-\frac{\delta^{ij}}{3}\,r^2\biggr)\theta(r-\epsilon)
+ \frac{4\,\pi}{15}\,\delta^3(r)\,(\gamma+\ln\epsilon)\,
\biggl(\partial^i\,\partial^j-\frac{\delta^{ij}}{3}\,\partial^2\biggr)\biggr]\,f(\bfr)\,,\nonumber
\end{align}
where $\gamma$ is the Euler constant.
The effective Hamiltonian $H_H$ represents the anomalous magnetic moment (amm) correction
to the Douglas-Kroll $m\alpha^6$ operators and is given by
\cite{pachucki:06:prl:he}
\begin{equation}
H_H = \sum_{i=1}^{17} H_i\,,
\end{equation}
where the $H_i$ are presented in Table~\ref{table1}. The last term of ${\cal E}^{(7)}_{\rm first}$ 
in Eq. (\ref{16}), the Hamiltonian $H^{(7)}_{\rm fs,amm}$ is the $m\alpha^7$ amm correction to the
Breit-Pauli Hamiltonian, see Eq. (\ref{HSD}).

The third part of $\cE^{(7)}$ is given by the second-order matrix elements
of the form
\cite{pachucki:06:prl:he}
\begin{align}
\cE^{(7)}_{\rm sec} =  &\
  2\lbr H^{(4)}_{\rm fs}\frac{1}{(E_0-H_0)'}H^{(5)}_{\rm nlog}\rbr
  + 2\lbr \biggl[ H^{(4)}_{\rm fs}+H^{(4)}_{\rm nfs}\biggr]
 \frac{1}{(E_0-H_0)'}H^{(5)}_{\rm fs}\rbr \,,
\end{align}
where $H^{(5)}_{\rm nlog}$ is the effective Hamiltonian responsible for the
nonlogarithmic $m\alpha^5$ correction to the energy
\begin{align}
H^{(5)}_{\rm nlog} = - \frac{7}{6\,\pi\,r^3}
+ \frac{38 Z}{45}\, \left[\delta^3(r_1)+\delta^3(r_2)\right] \,.
\end{align}
$H^{(4)}_{\rm nfs}$ is the spin-independent part of the Breit-Pauli
Hamiltonian given by Eq.~(\ref{nfs}),
and $H^{(5)}_{\rm fs}$ is the $m\alpha^5$ amm correction to $H^{(4)}_{\rm fs}$, see Eq. (\ref{HSD}).

The fourth part of $\cE^{(7)}$ is the contribution induced by the
emission and reabsorption of virtual photons of low energy.
It is denoted as $\cE_{L}^{(7)}$ and interpreted as
the relativistic correction to the Bethe logarithm.
The expression for $\cE_{L}^{(7)}$ reads
\cite{pachucki:00:jpb}
\begin{align}
 \cE_{L}^{(7)} = &
 -\frac{2}{3\,\pi}\,
\delta\, \Biggl< (\bfp_1+\bfp_2)\,\cdot (H_0-E_0)
\ln\left[\frac{2(H_0-E_0)}{Z^2}\right] 
(\bfp_1+\bfp_2)\Biggr>
  \nonumber \\ &
+\frac{i\,Z^2}{3\,\pi}
\Biggl< \left(\frac{\bfr_1}{r_1^3}+\frac{\bfr_2}{r_2^3}\right) \times
     \frac{\bsigma_1+\bsigma_2}{2}
\ln\left[\frac{2(H_0-E_0)}{Z^2}\right]
\left(\frac{\bfr_1}{r_1^3}+\frac{\bfr_2}{r_2^3}\right) \Biggr>\,,
\label{bethe}
\end{align}
\noindent
where $\delta \lbr \ldots \rbr$ denotes the first-order perturbation of
the matrix element $\langle\ldots\rangle$ by $H^{(4)}_{\rm fs}$, implying
perturbations of the reference-state wave function, the reference-state
energy, and the electron Hamiltonian.

\section{Results for helium fine-structure}

Summary of the individual contributions to the fine-structure intervals
of helium is given in Table~\ref{tab:summary}. Numerical results are
presented for the large $\nu_{01}$ and the small $\nu_{12}$ intervals,
defined by
\begin{eqnarray}
\nu_{01} &=& \bigl[ E(2^3P_0)-E(2^3P_1)\bigr]/h \\
\nu_{12} &=& \bigl[ E(2^3P_1)-E(2^3P_2)\bigr]/h\,.
\end{eqnarray}
We note that the style of
breaking the total result into separate entries used in
Table~\ref{tab:summary} differs from that used in the summary tables of
the previous papers by Pachucki {\em et al.} \cite{pachucki:06:prl:he,pachucki:09:hefs}.
In particular, the lower-order terms listed in Table III of
Ref.~\cite{pachucki:06:prl:he} and in Table~II of Ref.~\cite{pachucki:09:hefs}
contained contributions of higher orders, whereas in the present work
the entries in Table~\ref{tab:summary} contain only the contributions of the
order specified.

\begin{table*}
\caption{
Summary of individual contributions to the fine-structure intervals in helium,
in kHz. The parameters \cite{mohr:08:rmp} are
$\alpha^{-1} = 137.035\,999\,679(94)$, $cR_{\infty}
= 3\,289\,841\,960\,361(22)$~kHz, and $m/M = 1.370\,933\,555\,70 \times 10^{-4}$.
The label $(+m/M)$ indicates that the corresponding entry comprises both
the non-recoil and recoil contributions of the specified order in $\alpha$.
\label{tab:summary}
}
\begin{center}
  \begin{tabular}{l...}
\br
 Term &
\multicolumn{1}{c}{$\nu_{01}$}  &
        \multicolumn{1}{c}{$\nu_{12}$}  &  \multicolumn{1}{c}{$\nu_{02}$}\\
    \hline\\[-5pt]
$m\al^4(+m/M)$           &   29\,563\,765x.45        &        2\,320\,241x.43 &  \\[0.1cm]
$m\al^5(+m/M)$           &        54\,704x.04        &         -22\,545x.00 &  \\[0.1cm]
$m\al^6$                 &        -1\,607x.52(2)     &          -6\,506x.43 &  \\[0.1cm]
$m\al^6m/M$              &             -9x.96        &              9x.15 &  \\[0.1cm]
$m\al^7 \log(Z\al)$      &             81x.43        &             -5x.87 &  \\[0.1cm]
$m\al^7$, nlog           &             18x.86        &            -14x.38 &  \\[0.1cm]
$m\al^8$                 &           \pm1x.7         &           \pm1x.7  &  \\[0.1cm]
Total theory             &   29\,616\,952x.29 \pm1.7 & 2\,291\,178x.91\pm1.7 &       31\,908\,131x.20 \pm1.7 \\[0.1cm]
Experiment               &   29\,616\,951x.66(70)^a  &        2\,291\,177x.53(35)^d &31\,908\,131x.25(30)^f\\[0.1cm]
                         &   29\,616\,952x.7(10)^b   &        2\,291\,175x.59(51)^a &31\,908\,126x.78(94)^a\\[0.1cm]
                         &   29\,616\,950x.9(9)^c    &        2\,291\,175x.9(10)^e&\\
\br
\end{tabular}
\end{center}
$^a$ Ref.~\cite{zelevinsky:05},
$^b$ Ref.~\cite{giusfredi:05},
$^c$ Ref.~\cite{george:01},
$^d$ Ref.~\cite{borbely:09},
$^e$ Ref.~\cite{castillega:00},
$^f$ Ref.~\cite{shiner:10}.
\end{table*}

A term-by-term comparison with the independent calculation by Drake
\cite{drake:02:cjp} was performed in Ref. \cite{pach:10}. We observe good agreement
between the two calculations for the lower-order terms, namely, for the
$m\alpha^4$, $m\alpha^5$, and $m\alpha^6$
corrections. However, for the recoil correction to order $m\alpha^6$, our
results differ from those of Drake by about 0.5~kHz for both intervals. The
reason for this disagreement seems to be different for the large and the small
intervals. For the large interval, the deviation is due to the recoil operator
part, whereas for the small interval, it is mainly due to the mass
polarization part (see discussion in Ref.~\cite{pachucki:09:hefs}).

Our present estimates of the uncalculated higher-order effects for helium
are larger than those in the previous studies
\cite{pachucki:02:jpb:a,drake:02:cjp}. The previous estimates were
significantly less than 1~kHz. They  were based on 
logarithmic contributions to order $m\alpha^8$ corresponding to
the hydrogen fine structure. However, a larger contribution might 
originate from the nonlogarithmic relativistic corrections. 
So our present estimate is obtained 
by multiplying the $m\alpha^6$ contribution for the $\nu_{02} = \nu_{01}+\nu_{12}$ interval
by the factor of $(\Za)^2$, which yields a conservative
estimate of $\pm$1.7~kHz for all $\nu_{01}$, $\nu_{12}$, and $\nu_{02}$ intervals.
All nuclear structure effects are completely negligible at the current precision level.
The finite nuclear size correction
is estimated to yield 18 Hz for $\nu_{01}$ and 6 Hz for $\nu_{12}$.

Our result for the $\nu_{01}$ interval of helium agrees well with all recent experimental
values \cite{zelevinsky:05,giusfredi:05,george:01}. For the $\nu_{12}$ interval, 
theoretical result is by about $2\sigma$ larger than the values obtained in
Refs.~\cite{zelevinsky:05,castillega:00} but
in agreement with the latest measurement by Hessels and coworkers~\cite{borbely:09}.
Our theoretical prediction for the  $\nu_{02}$ interval is in excellent agreement
with the very recent measurement of Smiciklas and Shiner \cite{shiner:10}.
Comparison with this experimental result 
suggests that the higher-order contribution might in fact be much smaller than our
conservative estimate. 
This means that, if an independent measurement on Li$^+$ or Be$^{2+}$ confirms the
smallness of the $m\alpha^8$ terms,  
the helium determination of $\alpha$ will be significantly improved.
The measurement should be performed for the $2^3P_0 - 2^3P_2$ transition, 
since it is not affected by the singlet-triplet mixing effects, which strongly
depend on $Z$. 

In summary, the theory of the fine structure of helium and light helium-like
ions is now complete up to orders $m\alpha^7$ and $\alpha^6m^2/M$.
The theoretical predictions agree with the
latest experimental results for helium, as well as with most of the experimental
data for light helium-like ions. A combination of the theoretical and experimental results
\cite{shiner:10} for the $2^3P_0-2^3P_2$ interval in helium yields an independent determination of
the fine structure constant $\alpha$
\begin{equation}  \label{alpha}
\alpha^{-1} = 137.035\,999\,55(64)(4)(368)\,, 
\end{equation}
where the first error is the experimental uncertainty, 
the second one is the numerical uncertainty, and the third comes from the
estimate of the $m\,\alpha^8$ term ($\pm 1.7$ kHz). The result (\ref{alpha}) is
accurate to 27~ppb and in agreement with the recent value obtained from the electron
$g$ factor \cite{hanneke:08}. 

\ack
Support by NIST through Precision Measurement Grant PMG 60NANB7D6153
is gratefully acknowledged.


\section*{References}
\begin{thebibliography}{10}

\bibitem{schwartz:64}
Schwartz C 1964 Phys. Rev. {\bf 134} A1181

\bibitem{douglas:74}
Douglas M and Kroll N 1974 Ann. Phys. (NY) {\bf 82} 89

\bibitem{hambro:72}
Hambro L 1972 Phys. Rev. A {\bf 5} 2027

\bibitem{hambro:72:a}
Hambro L 1972 Phys. Rev. A {\bf 6} 865

\bibitem{hambro:73}
Hambro L 1973 Phys. Rev. A {\bf 7} 479

\bibitem{lewis:78}
Lewis M L and Serafino P H 1978 Phys. Rev. A {\bf 18} 867

\bibitem{zelevinsky:05}
Zelevinsky T, Farkas D and Gabrielse G 2005 Phys. Rev. Lett. {\bf 95} 203001

\bibitem{borbely:09}
Borbely J S, George M C, Lombardi L D, Weel M, Fitzakerley D W and Hessels E A 2009 Phys. Rev. A {\bf 79} 0605030(R)

\bibitem{shiner:10}
Smiciklas M and Shiner D 2010 Phys. Rev. Lett. {\bf 105} 123001 

\bibitem{yan:95:prl}
Yan Z -C and Drake G W F 1995 Phys. Rev. Lett. {\bf 74} 4791 

\bibitem{zhang:96:a}
Zhang T 1996 Phys. Rev. A {\bf 54} 1252

\bibitem{zhang:96:b}
Zhang T 1996 Phys. Rev. A {\bf 53} 3896 

\bibitem{zhang:96:prl}
Zhang T, Yan Z -C and Drake G W F 1996 Phys. Rev. Lett. {\bf 77} 1715

\bibitem{zhang:97}
Zhang T 1997 Phys. Rev. A {\bf 56} 270

\bibitem{pachucki:99:jpb}
Pachucki K 1999 J. Phys. B {\bf 32} 137

\bibitem{pachucki:00:jpb}
Pachucki K and Sapirstein J 2000 J. Phys. B {\bf 33} 5297

\bibitem{pachucki:02:jpb:a}
Pachucki K and Sapirstein J 2002 J. Phys. B {\bf 35} 1783

\bibitem{drake:02:cjp}
Drake G W F 2002 Can. J. Phys. {\bf 80} 1195

\bibitem{pachucki:03:jpb}
Pachucki K and Sapirstein J 2003 J. Phys. B {\bf 36} 803

\bibitem{pachucki:06:prl:he}
Pachucki K 2006 Phys. Rev. Lett. {\bf 97} 013002

\bibitem{pachucki:09:hefs}
Pachucki K and Yerokhin V A 2009 Phys. Rev. A {\bf 79} 062516;
{\em ibid.} 2009 {\bf 80} 019902(E); {\em ibid.} 2010 {\bf 81}  039903(E)

\bibitem{pachucki:10:hefs}
Pachucki K and Yerokhin V A 2010 Phys. Rev. Lett. {\bf 104} 070403

\bibitem{pach:10}
Pachucki K and Yerokhin V A 2010  Can. J. Phys. {\em in print}

\bibitem{kinoshita:07}
Aoyama T, Hayakawa M, Kinoshita T and Nio M 2007 Phys. Rev. Lett. {\bf 99} 110406

\bibitem{mohr:08:rmp}
Mohr P J, Taylor B N and Newell D B 2008 Rev. Mod. Phys. {\bf 80} 633

\bibitem{giusfredi:05}
Giusfredi G, Pastor P C, Natale P D, Mazzotti D, de~Mauro C, Fallani L,
Hagel G, Krachmalnicoff V and Inguscio M 2005 Can. J. Phys. {\bf 83} 301 

\bibitem{george:01}
George M C, Lombardi L D and Hessels E A 2001 Phys. Rev. Lett. {\bf 87} 173002

\bibitem{castillega:00}
Castillega J, Livingston D, Sanders A and Shiner D 2000 Phys. Rev. Lett. {\bf 84} 4321

\bibitem{hanneke:08}
Hanneke D, Fogwell S and Gabrielse G 2008 Phys. Rev. Lett. {\bf 100} 120801

\end{thebibliography}
\end{document}